\documentclass[runningheads]{llncs}

\usepackage{amsmath,amssymb,amsfonts}
\usepackage{algorithmic}
\usepackage{graphicx}
\usepackage{textcomp}
\usepackage{xcolor}
\usepackage{adjustbox}
\usepackage[caption=false]{subfig} 

\usepackage[multiple]{footmisc}
\usepackage{makecell}
\usepackage[hidelinks]{hyperref}
\usepackage{booktabs}

\newcommand{\repeatthanks}{\textsuperscript{\thefootnote}}

\begin{document}

\title{A Qualitative Evaluation of User Preference for Link-based vs. Text-based Recommendations of Wikipedia Articles}
\titlerunning{A Qualitative Evaluation of Wikipedia Recommendations}


\author{
Malte Ostendorff\inst{1}\thanks{Both authors contributed equally to this research.}
\and
 Corinna Breitinger\repeatthanks\inst{2}
 \and
 Bela Gipp\inst{2}
 }
 \authorrunning{M. Ostendorff et al.}
 \institute{Open Legal Data, Berlin, Germany \email{mo@openlegaldata.io}
 \and
 University of Wuppertal, Wuppertal, Germany
 \email{lastname@uni-wuppertal.de}
 }

\maketitle

\begin{abstract}

Literature recommendation systems (LRS) assist readers in the discovery of relevant content from the overwhelming amount of literature available. 
Despite the widespread adoption of LRS, there is a lack of research on the user-perceived recommendation characteristics for fundamentally different approaches to content-based literature recommendation.
To complement existing quantitative studies on literature recommendation, we present qualitative study results that report on users' perceptions for two contrasting recommendation classes:
(1)~link-based recommendation represented by the Co-Citation Proximity (CPA) approach, and 
(2)~text-based recommendation represented by Lucene's MoreLikeThis (MLT) algorithm. 
The empirical data analyzed in our study with twenty users and a diverse set of 40 Wikipedia articles indicate a noticeable difference between text- and link-based recommendation generation approaches along several key dimensions.
The text-based MLT method receives higher satisfaction ratings in terms of user-perceived similarity of recommended articles. In contrast, the CPA approach receives higher satisfaction scores in terms of diversity and serendipity of recommendations. 
We conclude that users of literature recommendation systems can benefit most from hybrid approaches that combine both link- and text-based approaches, where the user's information needs and preferences should control the weighting for the approaches used. The optimal weighting of multiple approaches used in a hybrid recommendation system is highly dependent on a user's shifting needs.

\keywords{Information retrieval \and recommender systems \and human factors \and recommender evaluation \and Wikipedia \and empirical studies.}
\end{abstract}

\section{Introduction}

The increasing volume of online literature has made recommendation systems an indispensable tool for readers. 
Over 50 million scientific publications are in circulation today \cite{Jinha2010}, and approx. 3 million new publications are added annually \cite{Johnson2018}. 
Encyclopedias such as Wikipedia are also subject to constant growth~\cite{SizeOfWikipedia}.
In the last decade, various recommendation approaches have been proposed for the literature recommendation use case.
In a review of 185 publications, 96 different approaches for literature recommendation were identified. 
The majority of approaches (55\%) continued to make use of content-based (CB) methods \cite{Beel2015}. Only 18\% of the surveyed recommendation approaches relied on collaborative filtering, and another 16\% made use of graph-based recommendation methods, i.e. the analysis of citation networks, author, or venue networks \cite{Beel2015}. 
A question that remains largely unexplored in today's literature is if these fundamentally different classes of recommendation algorithms are perceived differently by users. If a noticeable difference can be observed among users, across what dimensions do the end-users of such recommendation algorithms perceive that the approaches differ for a given recommendation use case?
The majority of studies dedicated to evaluating LRS make use of offline evaluations using statistical accuracy metrics or error metrics without gathering any qualitative data from users in the wild \cite{Beel2016b}. 
More recently, additional metrics have been proposed to measure more dimensions of user-perceived quality for recommendations, e.g. novelty \cite{Mendoza2020,Gravino2019}, diversity \cite{Kunaver2017,Yu2009}, serendipity \cite{Kaminskas2017,Kaminskas2014,DeGemmis2015,Ge2010}, and overall satisfaction \cite{Maksai2015,Joachims2005,Zhao2018}.
However, empirical user studies examining the perceived satisfaction with recommendations generated by different approaches remain rare. 
Given the emerging consensus on the importance of evaluating LRS from a user-centric perspective beyond accuracy alone \cite{Ge2010}, we identify a need for research to examine the user-perception of fundamentally different recommendation classes. 

In this paper, we perform a qualitative study to examine user-perceived differences and thus highlight the benefits and drawbacks of two contrasting LRS applied to Wikipedia articles. 
We examine a text-based recommendation generation approach, represented by Lucene's MLT \cite{LuceneMLT}, and contrast this with a link-based approach, as implemented in the Citolytics recommendation engine \cite{Schwarzer2017}, which uses co-citation proximity analysis (CPA) as a similarity measure \cite{Gipp2009}. 
Our study seeks to answer the following three research questions:

\begin{itemize}
    \item RQ1: Is there a measurable difference in users' perception of the link-based approach compared to the text-based approach? If so, what difference do users perceive? 
    \item RQ2: Do the approaches address different user information needs? If so, which user needs are best addressed by which approach?
    \item RQ3: Does one approach show better performance for certain topical categories or article characteristics?
\end{itemize}

Finally, we discuss how the evaluated recommendation approaches could be adapted in a hybrid system depending on the information needs of a user. 


\section{Background}\label{sec:background}

\subsection{Link-based Similarity Measure} \label{ssec:link-based}

Co-citation proximity analysis (CPA; \cite{Gipp2009}) determines the similarity among articles by comparing the patterns of shared citations or links to other works within the full text of a document. 
The underlying concept of CPA originates from the Library Science field, where it takes inspiration from the co-citation (CoCit) measure introduced by Small \cite{Small1973}. 
Beyond co-citation, CPA additionally takes into account the positioning of links to determine the similarity of documents. 
When the links of co-linked articles appear in close proximity within the linking article, the co-linked articles are assumed to be more strongly related. 

Schwarzer et al. \cite{Schwarzer2017} applied the concept of CPA to the outgoing links contained in Wikipedia articles. 
They introduced Citolytics as the first link-based recommendation system using the CPA measure and applied it to the Wikipedia article recommendation use case.
To quantify the degree of relatedness of co-linked articles, CPA assigns a numeric value, the Co-Citation Proximity Index (CPI), to each pair of articles co-linked in one or more linking articles. 
Schwarzer et al. \cite{Schwarzer2016} derived a general-purpose CPI that is independent of the structural elements of academic papers, e.g., sections or journals, which were proposed in the original CPA concept.
The general-purpose CPI is, therefore, more suitable for links found in the Wikipedia corpus. 
We define the co-link proximity $\delta_j(a,b)$ as the number of words between the links to article $a$ and $b$ in the article $j$. 
Equation~\ref{eq:cpi} shows the CPI for article $a$ being a recommendation for article $b$. 

\begin{equation}
\label{eq:cpi}
\text{CPI}(a,b)=\sum^{|D|}_{j=1}\delta_j(a,b)^{-\alpha}*log(\frac{|D|-n_a+0.5}{n_a+0.5})
\end{equation}

The CPI consists of two components: 
First, the general co-link proximity of $a$ and $b$ which is the sum of all marker proximities $\delta_j(a,b)$ over all articles in the corpus $D$.
The parameter $\alpha$ defines the non-linear weighting of the proximity $\delta$. 
In general, co-links in close proximity should result in a higher CPI than co-links further apart. 
Thus, $\alpha$ must be greater or equal to zero. 
Moreover, the higher $\alpha$, the closer the co-link proximity must be to influence the final CPI score. 
Prior to the user study, we conducted an offline evaluation similar to \cite{Schwarzer2016} and found that CPA achieves the best results with $\alpha=0.9$. 

The second component of CPI is a factor that defines the specificity of article $a$ based on its in-links $n_a$. 
This factor is inspired by the Inverse Document Frequency of TF-IDF, whereby we adapted the weighting schema from Okapi BM25 \cite{SparckJones2000}.
Hence, we refer to the factor as Inverse Link Frequency (ILF). 
We introduced ILF to counteract the tendency of CPA to recommend more general Wikipedia articles, which we discovered in a manual analysis of CPA recommendations. 
ILF increases the recommendation specificity by penalizing articles with many in-links, which tend to cover broad topics.

\subsection{Text-based Similarity Measure}
\label{ssec:text-based}

The text based \textit{MoreLikeThis} (MLT) similarity measure from Elasticsearch \cite{ElasticSearch} (based on Apache Lucence) differs fundamentally from the CPA approach. 
Instead of links, MLT relies entirely on the terms present in the article text to determine similarity. 
Using a Vector Space Model \cite{Salton1975}, MLT represents articles as sparse vectors in a space where each dimension corresponds to a separate index term. 
Term Frequency-Inverse Document Frequency (TF-IDF) proposed by  \cite{Jones1973} defines the weight of these index terms.
Accordingly, MLT considers two articles similar the more terms they share and the more specific these terms are. 
Thus, MLT-based article recommendations are more likely to cover similar topics when the topic is defined by specific terms that do not occur in other topics. 
MLT's simplicity and ability to find similar articles has made it popular for websites. 
For instance, MLT is currently used by Wikipedia's MediaWiki software, as part of its \textit{CirrusSearch} extension \cite{CirrusSearch}, to recommend articles to its users.

\subsection{Related Work}

Previously, Schwarzer et al. \cite{Schwarzer2016} performed an offline evaluation using the English Wikipedia corpus. 
They examined the offline performance of two link-based approaches, namely CPA and the more coarse CoCit measure, in addition to the text-based MLT measure. 
Schwarzer et al. made use of two quasi-gold standards afforded by Wikipedia. 
First, they considered the manually curated ‘see also’ links found at the end of Wikipedia articles as a quasi-gold-standard, which they used to evaluate 779,716 articles. 
Second, they used historical Wikipedia clickstream data in an evaluation of an additional 2.57 million articles. 
The results of this large-scale offline evaluation showed that the more fine-grained CPA measure consistently outperformed CoCit. This finding has also been validated by the research community in other recommendation scenarios \cite{Knoth2017}, \cite{Liu2011}.

Interestingly, this offline evaluation indicated that MLT performed better in identifying articles featuring a more narrow topical similarity with their source article. In contrast, CPA was better suited for recommending a broader spectrum of related articles \cite{Schwarzer2016}. 
However, prior evaluations by Schwarzer et al., using both `See also' links (found at the bottom of Wikipedia articles) and clickstream data, were purely data-centric offline evaluations. 
This prohibits gaining in-depth and user-centric insights into the users' perceived usefulness of the recommendations shown.
For example, click-through rates are a misleading metric for article relevance because users will click on articles with sensational or surprising titles before realizing that the content is not valuable to them.
Accuracy and error metrics alone 
are not a reliable predictor of a user's perceived quality of recommendations \cite{Cremonesi2011}. 
Only user studies and online evaluations can reliably assess the effectiveness of real-world recommendation scenarios.


Knijnenburg et al. \cite{Knijnenburg2012} proposed a user-centered framework that explains how objective system aspects influence subjective user behavior. Their framework is extensively evaluated with four trials and two controlled experiments and attempts to shed light on the interactions of personal and situational characteristics. Additionally, they take into account system aspects to explain the perceived user experience for movie recommendations.

Pu et al. \cite{Pu2011} developed a user-centric evaluation framework termed ResQue (Recommender system's quality of user experience) consisting of 32 questions and 15 constructs to define the essential qualities of an effective and satisfying recommender system, including the recommendation qualities of accuracy, novelty, and diversity.
Since their framework can be applied to article recommendations, we include several questions from the ResQue framework in our evaluation design (refer to Section~\ref{ssec:study-design} for details). 

Despite the large size and popularity of Wikipedia, the potential of this corpus for evaluating LRS has thus far not been exploited by the research community. 
To the best of our knowledge, no prior work, aside from the initial offline study \cite{Schwarzer2016} and the work by  \cite{Molloy2020}, has made use of the Wikipedia corpus to evaluate the effectiveness of different recommendation approaches. 
The implications of studying the user-perceived recommendation effectiveness for Wikipedia articles may also be applicable to Wikimedia projects in a broader context, which tend to contain a high frequency of links. 

\section{Methodology}\label{sec:methodology}

This section describes our study methodology and the criteria for selecting the Wikipedia articles used in our study.
The Wikipedia encyclopedia is one of the most prominent open-access corpora among online reference literature.
As of June 2021, the English Wikipedia contains approximately 6.3 million articles \cite{SizeOfWikipedia}.
Its widespread use 
and accessibility motivated us to use the English Wikipedia to source the articles for our live user study. 
We consulted the same English Wikipedia corpus as in \cite{Schwarzer2016} with the pre-processing as in \cite{Schwarzer2017}. 

\subsection{Study Design}
\label{ssec:study-design}



Prior to our study, we created a sample of 40 seed articles covering a diverse spectrum of article types in Wikipedia. 
When selecting these seed articles, our aim was to achieve a diversity of topics, which nonetheless remained comprehensible to a general audience.
To ensure comprehensibility, we excluded topics that would require expert knowledge to judge the relevance of recommendations, e.g., articles on mathematical theorems. 
Moreover, the seed articles featured diverse article characteristics, such as article length and article quality\footnote{\label{fn:quality}We judge the article quality using Wikipedia's vital article policy \cite{WikipediaVitalArticles}.}.

We distinguished seed articles into four categories. 
First, according to their \textit{popularity} (measured by page views) into either niche or popular articles, and second, according to the content of the article into either \textit{generic}, i.e., reference articles typical of encyclopedias, or \textit{named entities}, i.e., politicians, celebrities, or locations. 
We choose popularity as a criterion because, on average, popular articles receive more in-links from other articles. 
Schwarzer et al. \cite{Schwarzer2016} found that the number of in-links affected the performance of the link-based CPA approach.
Moreover, we expect study participants to be more familiar with popular topics compared to niche articles. Therefore users will be better able to verbalize their spontaneous information needs when examining a topic.
The `article type' categories were chosen to study the effect that articles about named entities may have on MLT. 
Names of entities tend to be more unique than terms in articles on generic topics. Therefore, we expect that specific names may affect MLT's performance.
Likewise, due to the nature of Wikipedia articles linking to generic topics, they may appear in a broader context than links to named entities.
Thus, CPA's performance may also be affected. 
These considerations resulted in four article categories: 
(A) niche generic articles, 
(B) popular generic articles, 
(C) niche named entities, and 
(D) popular named entities. 
Table~\ref{tab:seed} shows these four categories and the 40 seed articles selected for recommendation generation.

To perform our qualitative evaluation of user-perceived recommendation effectiveness, we recruited 20 participants. 
Participants were students and doctoral researchers from several universities in Berlin and the University of Konstanz.
The average age of participants was 29 years. 
65\% of our participants said they spend more than an hour per month on Wikipedia, with the average being 4.6 hours spent on Wikipedia. 

Our study contained both qualitative and quantitative data collection components. The quantitative component was in the form of a written questionnaire.
This questionnaire asked participants about each recommendation set separately and elicited responses on a 5-point Likert scale. 
Some questions were tailored to gain insights on the research questions we defined for our study. The remainder of the questions adhered to the ResQue framework for user-centric evaluation \cite{Pu2011}.
The qualitative data component was designed as a semi-structured interview. 
The interview contained open-ended questions that encouraged participants to verbally compare and contrast the two recommendation sets. 
The participants were also asked to describe their perceived satisfaction.
Resulting from this mixed methods study design, we could use the findings from the qualitative interviews to interpret and validate the results from the quantitative questionnaires. All interviews were audio-recorded with the permission of our participants.

In the study, each participant was shown four Wikipedia articles, one at a time. 
For each article, two recommendation sets, each containing five recommended articles, were displayed. 
One set was generated using CPA, i.e., the link-based Citolytics implementation \cite{Schwarzer2017}, while the other was generated using the MLT algorithm. 
Each set of four Wikipedia articles was shown to a total of two participants to enable checking for the presence of inter-rater agreement. 
Participants were aware that recommendation sets had been generated using different approaches, but they did not know the names of the approaches or the method behind the recommendations. We alternated the placement of the recommendation sets to avoid the recognition of one approach over the other and forming a potential bias based on placement. 
The seed Wikipedia articles were shown to participants via a tablet or a laptop. 
The participants were asked to read and scroll through the full article so that the exploration of the article's content was a natural as possible.
We have made the complete questionnaire and the collected data publicly available on GitHub\footnote{\label{fn:github}\url{https://github.com/malteos/wikipedia-article-recommendations}}.

\begin{table*}[htp] 

\setlength{\tabcolsep}{3.5pt} 
\renewcommand{\arraystretch}{1.0} 

\caption{\label{tab:seed}Overview of seed articles selected for the study.}

\centering
\begin{adjustbox}{width=1.0\textwidth}
\begin{tabular}{clrcclr}

\toprule

\textbf{\#} 
& \textbf{Article (Quality\footref{fn:quality})}                                                
& \textbf{Words} 
&  
& \textbf{\# }
& \textbf{Article (Quality\footref{fn:quality})}                                              
& \textbf{Words} 

\\

\midrule

\textit{A}        & \textit{Niche generic topics}                                   &            &                               &     \textit{C }   & \textit{Niche named entities} &                                        \\
\rule{0pt}{3ex}

1         & Babylonian mathematics (B)                               & 3,825                           &  & 21        & Mainau (S)                                              & 567                          \\
2         & Water pollution in India (S)                               & 1,697                        &  & 22        & Lake Constance  (C)                                      & 7,079                          \\
3         & Transport in Greater Tokyo  (C)                            & 3,046                     &  & 23        & Spandau (C)                                              & 599                              \\
4         & History of United States cricket (S)                       & 3,610                           &  & 24        & Appenzell (C)                                            & 2,667                            \\
5         & Firefox for Android  (C)                                 & 4,821                          &  & 25        & Michael Müller (politician) (Stub)                         & 602                         \\
6         & Chocolate syrup (Stub)                                       & 391                   &  & 26        & Olympiastadion (Berlin) (C)                               & 3,360                          \\
7         & Freshwater snail   (C)                                    & 1757                           &  & 27        & Theo Albrecht   (S)                                      & 929                        \\
8         & Touring car racing  (S)                                   & 2550                            &  & 28        & ARD (broadcaster)   (S)                                  & 2,397                           \\
9         & Mudflat   (C)                                             & 787                           &  & 29        & Kaufland    (Stub)                                          & 680                           \\
10        & Philosophy of healthcare (B)                              & 3,804                           &  & 30        & Sylt Air  (Stub)                                            & 110                            \\

\rule{0pt}{4ex}   

\textit{B}         &  
\textit{Popular generic topics}
&                                      &  & \textit{D}         & 
\textit{Popular named entities}
&                                    \\

\rule{0pt}{3ex}   

11        & Fire  (C)                                                 & 4,297                  &  & 31        & Albert Einstein (GA)                                      & 15,071                     \\
12        & Basketball  (C)                                           & 11,172                  &  & 32        & Hillary Clinton   (FA)                                    & 28,645                    \\
13        & Mandarin Chinese  (C)                                     & 6.98                    &  & 33        & Brad Pitt  (FA)                                           & 9,955                   \\
14        & Cancer     (B)                                            & 16,300                &  & 34        & New York City   (B)                                      & 30,167                \\
15        & Vietnam War   (C)                                         & 32,847                &  & 35        & India (FA)                                                & 16,861                  \\
16        & Cat           (GA)                                         & 17,009             &  & 36        & Elon Musk  (C)                                           & 11,529                \\
17        & Earthquake      (C)                                       & 7,541                  &  & 37        & Google   (C)                                             & 16,216                   \\
18        & Submarine       (C)                                       & 11,968                    &  & 38        & Star Wars    (B)                                         & 16,046                     \\
19        & Rock music      (C)                                       & 19,833                &  & 39        & AC/DC      (FA)                                           & 10,442                   \\
20        & Wind power     (GA)                                        & 15,761                  &  & 40        & FIFA World Cup    (FA)                                    & 7,699        

\\
\bottomrule

\end{tabular}
\end{adjustbox}

\end{table*}

\section{Results}\label{sec:results}

In this section, we summarize and discuss the empirical data collected.
First, we present the primary findings, in which we provide answers to the three research questions specified in the Introduction, and illustrate them with participants' quotes. 
Second, we discuss secondary findings that arose from coding the participants' responses, which go beyond the research questions we set out to answer.

\subsection{Primary Findings}

\begin{figure}[ht]
\captionsetup[]{labelfont=bf}
  \centering
  \includegraphics[width=0.99\linewidth,trim=0.5cm 1.25cm 1.2cm 0cm, clip]{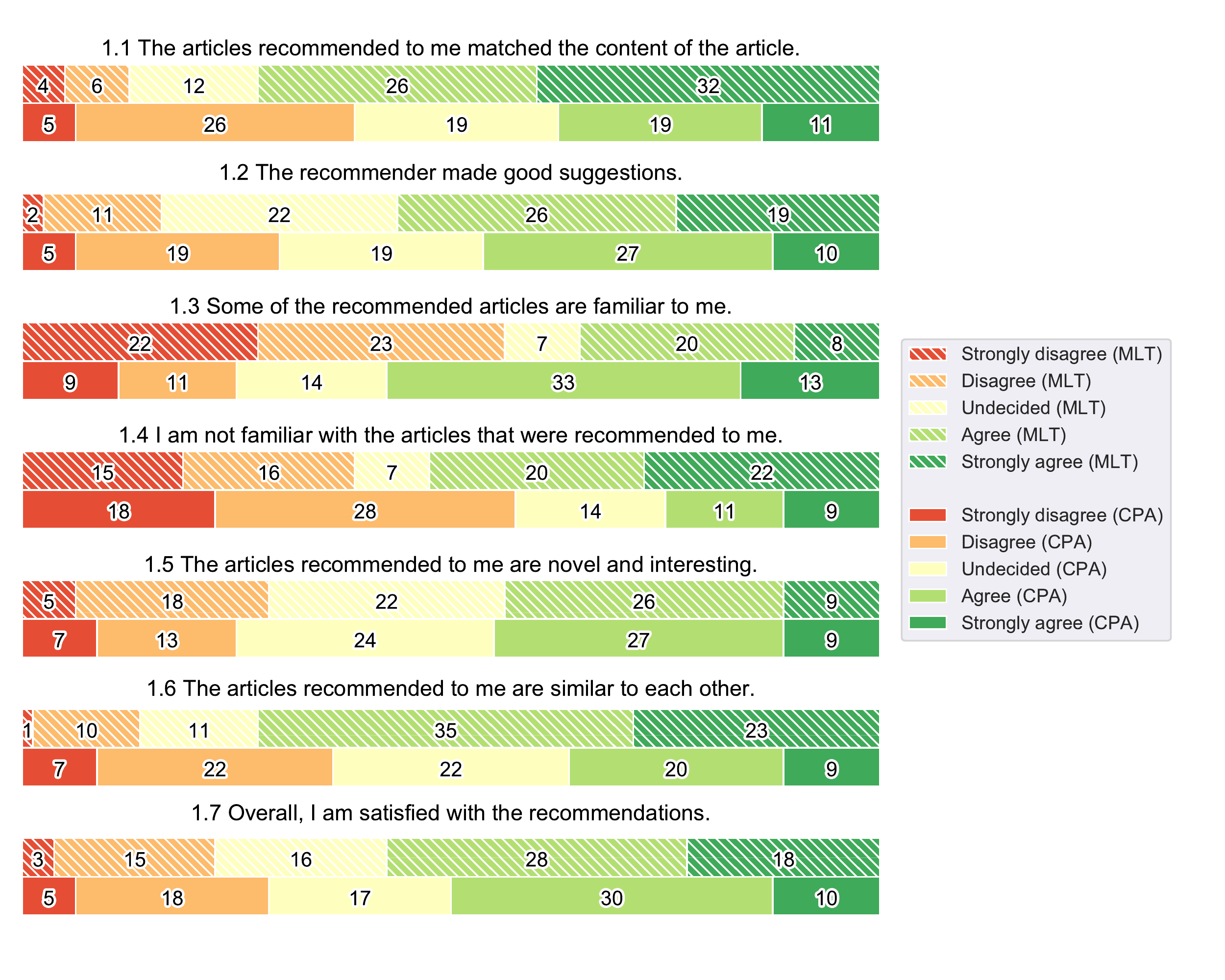}
  \caption{\label{fig:answers}
  Responses for MLT (dashed) and CPA (solid) on a 5-point Likert scale.}
\end{figure}

Our study found several differences in reader's perception of the link-based approach compared to the text-based approach. 
A notable difference could be identified especially in the perceived degree of `similarity' of the recommendations.
Participants were significantly more likely to agree with the statement `the recommendations are more similar to each other' (see 1.6 in \figurename~\ref{fig:answers}) for the MLT approach. 73\% of responses `agreed' or `strongly agreed' (58 out of 80 responses) with this statement, compared to only 36\% of the responses for the CPA approach (29 out of 80). Keep in mind each of the 40 seed articles was examined by two participants resulting in 80 responses in total.
A question about whether the articles being recommended `matched with the content' of the source article (see 1.1) was answered with a similar preference, with a significantly higher portion of the responses indicating `strongly agree' or `agree' for the MLT approach (73\%) and only 38\% of responses choosing the same response for the CPA approach.

Overall, users perceived recommendations of CPA as more familiar (see 1.3). They felt less familiar (1.4) with the recommendations made by MLT.
We found that this difference was observed by nearly all participants and can be attributed to how MLT considers textual similarity.
In general, MLT focuses on overlapping terms, while CPA utilizes the co-occurrence of links. 
The quantitative results in \cite{Schwarzer2016} already suggested that this leads to diverging recommendations.



\begin{figure}[htp] 
     \centering
     \subfloat[\label{fig:diverse-similar}  Diverse or similar.]{
        \includegraphics[width=0.4\linewidth,trim=0.5cm 0.5cm 0.5cm 0.3cm, clip]{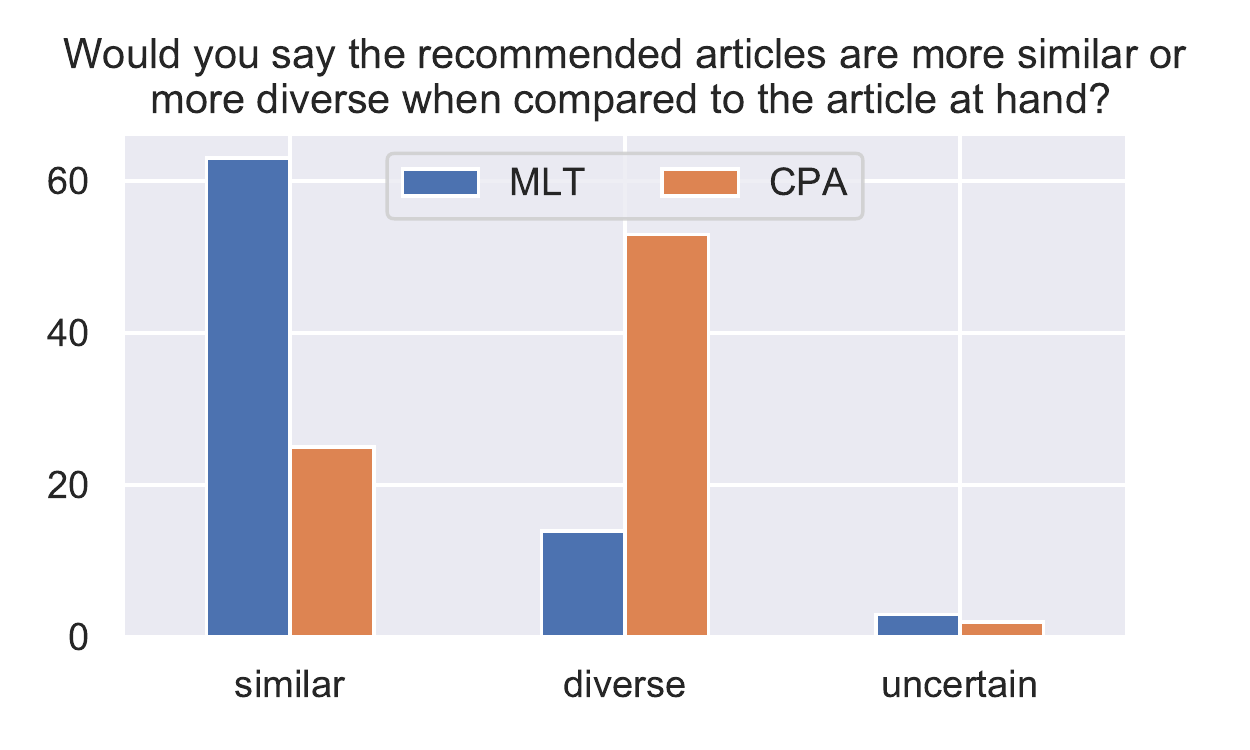}
     }
     \subfloat[\label{fig:familiarity} Familiarity with article topic.]{
        \includegraphics[width=.5\linewidth,trim=0.5cm 0.5cm 0.5cm 0.3cm, clip]{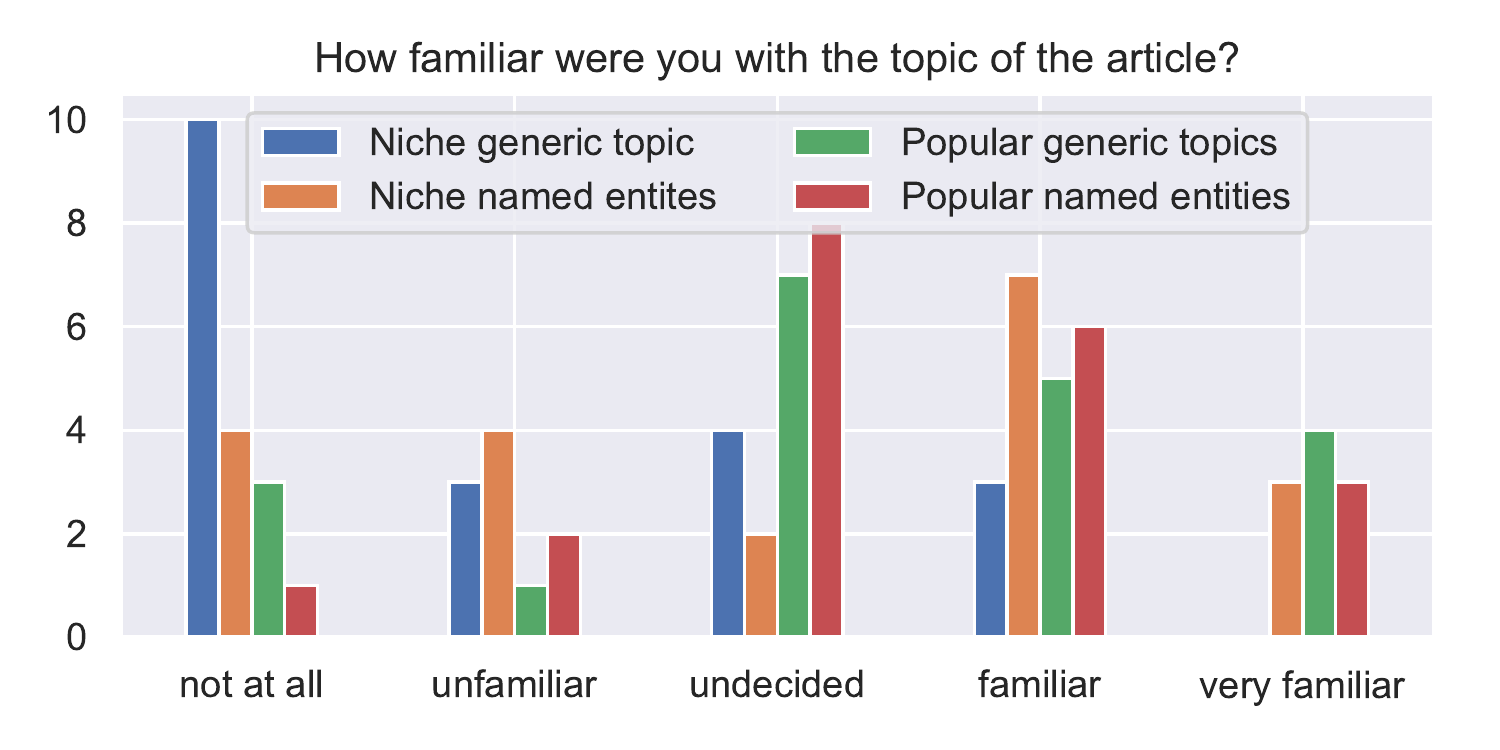}
     }
        \caption{Quantitative answers}
        \label{fig:three graphs}
\end{figure}

\paragraph{Perceived Difference between CPA and MLT.}

The participants observed that the methodological difference between the approaches affected their recommendations.
In the questionnaire, participants expressed 48 times that the articles recommended by CPA are more diverse, i.e., less similar, compared to the seed article (\figurename~\ref{fig:diverse-similar}).
MLT’s recommendations were found to be diverse only 13 times.
Regarding the similarity of recommendations, the outcome is the opposite.

The participants' answers also indicate the difference between MLT's and CPA's recommendations.
Participant P20 explained that \textit{``approach A [CPA] is more an overview of things and approach B [MLT] is focusing on concrete data or issues and regional areas''}.
CPA providing an ``overview of things'' is not favorable for all participants as they describe different information needs.
For example, participant P20 prefers MLT's recommendation since \textit{``it is better to focus on the details''}.
Some participants attributed the recommendations' similarity (or diversity) to terms co-occurring in the title of the seed and recommended articles. 
For instance, participant P15 found MLT's recommendations for \textit{Star Wars} to be more similar because \textit{``Star Wars is always in the title [of MLT's recommendations]''}.
Participant P17 also assumes a direct connection between the seed and MLT's recommendations \textit{``I'd guess recommendations of A [MLT] are already contained as link in the source article''}.

Nonetheless, participants struggled to put the observed difference between MLT and CPA in words, although they noticed categorical differences in the recommendation sets.
Participant P19 said \textit{``I can see a difference but I don't know what the difference is''}.
Similarly, participant P20 found that \textit{``they [MLT and CPA] are both diverse to the same extent but within a different scope''}.

Concerning the overall relevancy of the recommendations, MLT outperformed CPA.
In total, the participants agreed or strongly agreed 45 times that MLT made `good suggestions' (see 1.2), whereas only 37 times the same was stated for CPA.
Similarly, the overall satisfaction was slightly higher for MLT (46 times agree or strongly agree) compared to CPA (40 times; see 1.7).


\paragraph{Information Need.}

The participants are also aware of relevancy depending on their individual information need. 
When asked about the `most relevant recommendation' the participants' answers contained the words `depends' or `depending' ten-times. 
Participant P15 states that \textit{``if I want a broader research I'd take B [CPA] but if I'd decide for more punctual research I would take A [MLT] because it is more likely to be around submarine and because in B [CPA] I also get background information''.}
Similarly, participant P13 would click on a recommendation as follows: \textit{``if you're looking for a specific class/type of snails then this [MLT] could be one, but if you're just looking to get an overview of aquatic animals, then probably you would click on the other approach [CPA]''}.
In summary, the participants agreed on CPA providing `background information' that is useful to `get an overview of a topic', while MLT's recommendations were perceived as `more specific' and having a `direct connection' to the seed article.


For articles on science and technology, the most commonly expressed information needs were understanding how a technology works or looking up a definition. For articles about individuals, participants expressed the need to find dates relating to an individual and to understand their contributions to society. 
For `niche' topics, users were slightly more likely to state the desire to discover sub-categories on a topic, which implies wishing to move from a broader overview to a more fine-grained and in-depth examination of the topic.  

The subjectiveness 
is also reflected by the inter-rater agreement.
The participants who reviewed the same articles had a Cohen's kappa of $\kappa=0.14$ on average, which corresponds to slight agreement.
The inter-rater agreement increases to a ``fair agreement'' (see \cite{fairkappa})  when we move from a 5-point to a 3-point Likert scale, i.e., possible answers are `agree', `undecided', or `disagree'.
Low agreement indicates that the perception of recommendation highly depends on the individual's prior knowledge and information needs.


\paragraph{Article Characteristics.}

%
The article `types', which we defined as described in the methodology section according to article popularity, length and breadth into the four categories `popular generic', `niche generic', `popular named entities', and `niche named entities' had no observable impact on user's preference for one recommendation approach over the other.

However, we found that the user-expressed information need, for example, the desire to identify related articles that were either more broadly related or were more specialized, did have a measurable impact on the user’s preference for the recommendation method.
For \textit{popular generic articles} on science and technology, e.g., the article on wind power, the most frequently expressed information needs were understanding how a technology works or looking up definitions.

For articles in the categories `popular generic' and `niche generic', we could observe that the information needs expressed by our readers were more \textit{broad}. For example, they wanted to find definitions for the topic at hand, more general information to understand a topic in its wider context, or examples of sub-categories on a topic. There was no observable difference between the specified categories of information need for `popular' vs. `niche' generic articles.

Resulting from our initial classification of the 40 Wikipedia articles selected, the empirical questionnaire data showed that `niche' entities were on average more familiar to the participants than we initially expected (\figurename~\ref{fig:familiarity}).
This was especially the case for niche named entities, many of which were rated as being familiar to the participants. 
The reason for this may be that our participants were from Germany and were thus familiar with many of these articles, despite the articles reporting on regional German topics, e.g., Spandau, Mainau. On the other hand, users rated niche generic topics as being less familiar, which was in line with what we expected.

Furthermore, both popular generic topics and popular named entities were less often classified as `unfamiliar' by the participants than they were classified as `neutral or familiar'. 
Lastly, one notable finding is that Wikipedia listings, e.g., \textit{List of rock genres} or \textit{List of supermarket chains in Germany}, were found to be the most relevant recommendations in some cases.
The CPA implementation \cite{Schwarzer2017} intentionally excludes Wikipedia listings from its recommendation sets.
Thus, the implementation needs to be revised accordingly.


\subsection{Secondary Findings}

\begin{figure}[htp] 
  \centering
  \includegraphics[width=0.99\linewidth,trim=0 0.0cm 2.5cm 0cm, clip]{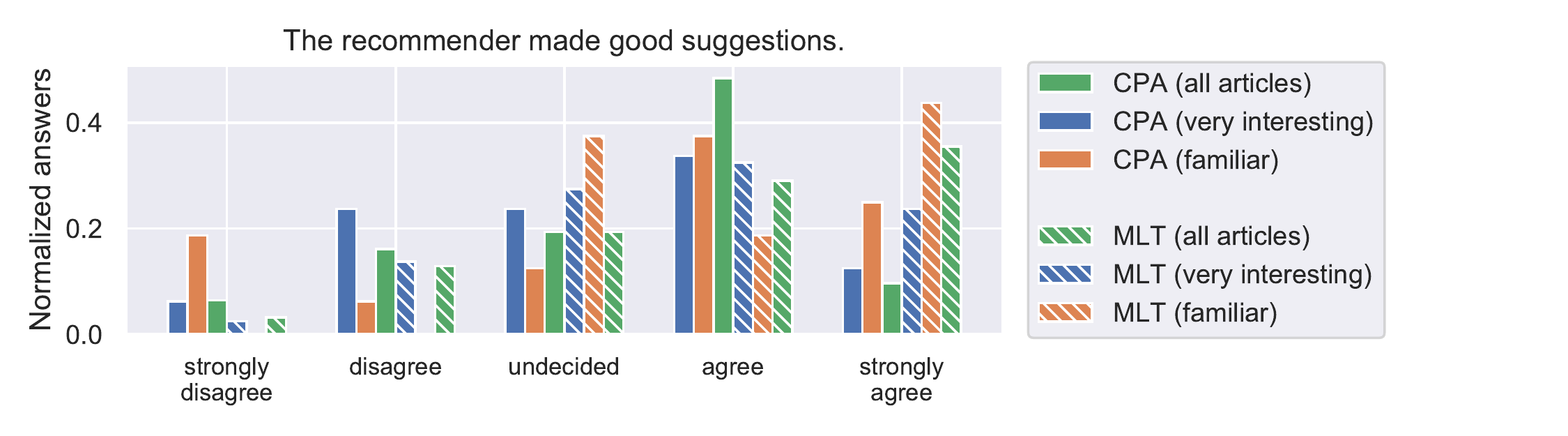}
  \caption{\label{fig:satisfaction}
  User's satisfaction depending on interest and familiarity, i.e., for all articles or only the articles which are very interesting or familiar to the user.}
\end{figure}


\paragraph{Effect of User's Interest on Recommendations.}

\figurename~\ref{fig:satisfaction} shows that MLT outperforms CPA in recommendation satisfaction if participants `strongly agreed' with the article topic being (a) interesting or (b) being at least `familiar' to them.
This is likely the case because users are more versed in judging the relevance of the text-based recommendations of the MLT approach if they already have more in-depth knowledge of a topic. 
For example, one participant observed regarding CPA's recommendations of \textit{Renewable energy} for \textit{Wind power} as the seed article: \textit{``Renewable energy is least relevant because everybody knows something about it [Renewable energy]''}. 
Sinha and Swearingen \cite{Sinha2002} have already shown previous familiarity with an item as a confounding factor on a user `liking' a recommendation.
Interestingly, this trend was no longer observable in cases when participants only `agreed' but without strong conviction that the articles were interesting or familiar to them. 
In these cases, the MLT and CPA approaches were seen as more equal, with the CPA approach taking a slight lead. 


\paragraph{User-based Preferences.}

Our findings confirm the subjectiveness of recommendation performance, since we observe user-based preferences.
For instance, participant P2 only agreed or strongly agreed for CPA on MC 1.2 `The recommender made good suggestions' and 1.7 `Overall, I am satisfied with the recommendations' but never gave the same answers for MLT.
However, participant P3 showed the opposite preference, i.e., only MLT made good suggestions according to P3.
The remaining participants had more balanced preferences. 
In terms of MC 1.2 and 1.7, nine participants had a tendency to prefer MLT, while six participants preferred CPA, and five participants did not show any particular preference for one of the two recommendation approaches.

\paragraph{Perception of Novelty \& Serendipity of Recommendations.}

To gain insights on the user-perceived novelty, we asked participants to rate the following statements: `I am not familiar with the articles that were recommended to me' and `The articles recommended to me are novel and interesting'.
While novelty determines how unknown recommended items are to a user, serendipity is a measure of the extent to which recommendations positively surprise their users~\cite{Ge2010,Kaminskas2017}.
For instance, participant P19 answered \textit{``approach A [MLT] shows me some topics connected with the article I read but with more special interest - they are about ``Healthcare'', and B [CPA] actually changes the whole topic. B [CPA] offers totally different topics.''} 
CPA's recommendations are generally found to be more serendipitous.
For the question regarding an `unexpected recommendation among the recommendation set' CPA received a yes answer 41 times, compared to only 23 times for MLT.
The perceived novelty also made participants click on recommendations.
Among others, participant P1 explained that \textit{``there are more [MLT] articles that I would personally click on, because they are new to me.''}
Similarly, participant P16 stated that they would \textit{``click first on Star Wars canon, because I don’t know what it is''}.



\paragraph{Trust \& (Missing) Explanations.}
%
Although the questionnaire is not designed to investigate the participants' trust in the recommendations, many answers addressed this topic. 
When users were asked for the relevancy of recommendations, some participants expressed there ``must be a connection'' between the article at hand and a recommendation and that they just ``do not know what is has to do with it''. 
Others were even interested in topically irrelevant recommendations.
For example, they expressed \textit{``it interests me why this is important to the article I am reading''.} 
Similarly, a participant said they might click on a recommendation \textit{``because I do not know what it has to do with [the seed article]''.} 
Such answers were more often found for CPA recommendations since they tend to be more broadly related than MLT's more narrow topical similarity. 
In some cases, there is no semantic relatedness. Yet, even then, participants often do not recognize a recommendation as irrelevant. Instead, they say it is their fault for not knowing how the recommendation is relevant to the seed. 
This behavior indicates a high level of trust from the participants placed in the recommender system.


\section{Discussion}

The experimental results demonstrate that MLT and CPA differ in their ability to satisfy specific user information needs. Furthermore, our study participants were capable of perceiving a systematic difference between the two approaches.

CPA was found to provide an `overview of things' with recommendations more likely to be unfamiliar to the participants and less likely to match with the content of the seed article.
In contrast, MLT was found to `focus on the details'. Participants also felt that MLT's recommendations matched the content of the seed article more often.
At the same time, participants perceive CPA's recommendations as more diverse, while MLT's recommendations are more similar to each other. 
So CPA and MLT, being conceptually different approaches and relying on different data sources, lead to unique differences in how the recommendations were perceived.
In terms of the overall satisfaction with recommendations, most participants expressed a preference for MLT over CPA.
MLT is based on TF-IDF and, therefore, its recommendations are centered around specific terms (e.g., P15: \textit{``Star Wars is always in the title''}).
In contrast, CPA relies on the co-occurrence of links.
According to CPA, two articles are considered related when they are mentioned in the same context.
Our results show that this leads to more distantly related recommendations, which do not necessarily share the same terminology.
Given that the participants experience the two recommendation approaches differently, a hybrid text- and link combination, depending on the context, is preferable (as demonstrated in \cite{Kanakia2019,Farber2020,Ostendorff2021}).

Moreover, the differently perceived recommendations show the shortcoming of the notion of similarity. 
Both approaches, CPA and MLT, were developed to retrieve semantically similar documents, which they indeed do~\cite{Gipp2009,Jones1973}.
However, their recommendations are `similar' within different scopes. 
A recommended article that provides an `overview' can be considered similar to the seed article. 
Equivalently, a `detailed' recommendation can also be similar to the seed but in a different context.
Our qualitative interview data could show how users perceive these two similarity measures differently.
These findings are aligned with \cite{Bar2011}, which found that text similarity inherits different dimensions.

We also found that either CPA's or MLT's recommendations are liked or disliked depending on the individual participant preferences.
Some participants even expressed a consistent preference for one method over the other.
However, a strict preference was the exception.
We could also not identify any direct relation between the user or article characteristics and the preference for one method.
At this point, more user data as in a user-based recommender system would be needed to tailor the recommendations to the user's profile.
Purely content-based approaches such as CPA and MLT lack this ability \cite{Jannach2010,Beel2015,Lenhart2016}.
The only option would be to allow users to select their preferred recommendation approach through the user interface depending on their information need.

The participants' answers also revealed a trust in the quality of the recommendations that was not always justified.
Participants would assume a connection between the seed article and the recommended article just because it was recommended by the system.
Instead of holding the recommender system accountable for non-relevant recommendations, participants found themselves responsible for not understanding a recommendations relevance.
To not disappoint this trust, recommender systems should provide explanations that help users understand why a particular item is recommended.
Also, explanations would help users to understand connections between seed and recommendations.
Explainable recommendations are a subject of active research \cite{Zhang2020,Kunkel2019}.
However, most research focuses on user-based approaches, while content-based approaches like CPA or MLT could also benefit from explanations.

Despite the insights of our qualitative study to elicit user's perceived differences in recommendation approach performance, the nature of our evaluation has several shortcomings. With 20 participants, the study is limited in size.
Consequently, our quantitative data points suggest a difference that is not statistically significant. 
Large-scale offline evaluations (e.g., \cite{Schwarzer2016}) are more likely to produce statistically significant results.
For this reason, and for not requiring participants, such offline evaluations are more commonly used in recommender system research~\cite{Beel2015}.
But offline evaluations only provide insights in terms of performance measures.
Our study shows that this can be an issue. 
When consulting only our quantitative data, one could assume that MLT and CPA are comparable in some aspects since their average scores are similarly high. 
The discrepancies between CPA and MLT only become evident when analyzing the written and oral explanations of live users.
This highlights that recommender system research should not purely rely on offline evaluations~\cite{Beel2016b}.

Moreover, we acknowledge that recommendations of Wikipedia articles differ from recommendations of other literature types.
It is thus uncertain whether our findings relating to encyclopedic recommendations in Wikipedia can be directly transferred to other domains.
Recent advancements in recommender system research are focused on neural-based approaches, which may lead to the belief that the examined methods, MLT and CPA, are dated.
This, however, is not the case since they are still used in practice, as Wikipedia's MLT deployment shows (Section~\ref{ssec:text-based}), and the intuition of CPA is the basis for neural approaches like in Virtual Citation Proximity \cite{Molloy2020}.

\section{Conclusion}

We elicit the user-perceived differences in performance for two well-known recommendation approaches.
With the text-based MLT and the link-based CPA, we evaluate complementary content-based recommender system implementations. 
In a study with 20 participants, we collect qualitative and quantitative feedback on recommendations for 40 diverse Wikipedia articles.

Our results show that users are generally more satisfied with the recommendations generated by text-based MLT, whereas CPA's recommendation are perceived as more novel and diverse.
The methodological difference of CPA and MLT, i.e., being based on either text or links, is reflected in their recommendations and noticed by the participants.
Depending on information needs or user-based preferences, this leads to one recommendation approach being preferred over the other.
Thus, we suggest combining both approaches in a hybrid system, since they both address different information needs.

As a result of the insights gained from our study, we plan to continue research on a hybrid approach tailored to the recommendation of literature, which accounts for diverse information needs.
Moreover, we will investigate how content-based features can be utilized to provide explanation such that users can understand why a certain item is recommend to them.
%
Lastly, we make our questionnaires, participants' answers, and code publicly available\footref{fn:github}. 

\subsubsection*{Acknowledgements}
We thank the anonymous reviewers, all participants, and especially Olha Yarikova for her assistance in the completion of the interviews.


\bibliographystyle{splncs04}
\bibliography{paper}

\end{document}